\documentclass[preprint,12pt]{elsarticle}

\usepackage{amssymb}

\journal{Physica E}

\begin{document}

\begin{frontmatter}



\title{Tip apex charging effects in tunneling spectroscopy}


\author{T. Kwapi\'nski\footnote[1]{e-mail address: tomasz.kwapinski@umcs.lublin.pl}
and M. Ja\l ochowski}

\address{
Institute of Physics, M. Curie-Sk\l odowska University, \\
20-031 Lublin, Poland}

\begin{abstract}
The influence of charged STM tip on the electron transport through
quantum states on a surface is studied both theoretically and
experimentally. The current and the differential conductance
calculations are carried out by means of the Green's function
technique and a tight-binding Hamiltonian.  It is shown that sharp
STM tip is extra occupied and this additional charge breaks the
conductance symmetry for positive and negative STM voltages. The
experiment on Ag islands with two  STM tips (blunt and sharp)
confirms our theoretical calculations.
\end{abstract}


\end{frontmatter}


\section{\label{sec10}Introduction}

The invention of scanning tunnelling microscope (STM) in the early
1980's \cite{111} revolutionized imaging processes of various
surfaces. This instrument allows us to study exact positions of
atoms (in conducting surface or in small nanostructures at the
surface) with atomic resolution. Moreover, the electrical
properties (tunnelling spectroscopy, STS) can be investigated as
well.

The STM resolution depends on the tip quality: the sharper tip the
better resolution can be achieved and only one atom at the end of
the tip is the ideal case. The last tip atom properties are
crucial for electron transport through the STM,
\cite{Pelz,Hofer01,Passoni,Hofer02}. Using such a sharp tip the
spectroscopy or topography studies of atomic objects and quantum
states on a surface are possible with the best resolution. In
order to obtain useful information on density of states of
investigated objects an appropriate treatment of STS results is
required. The most popular theoretical description of the STS
transport have originally been developed by Tersoff and Hamann
\cite{Tersoff}. Within the WKB approximation the tunnelling
current is expressed in terms of the surface and tip local density
of states (DOS) and the transmittance through the system. However,
the energy convolution between both density of states makes it
difficult to interpret STS results, e.g. \cite{Passoni,Li}. It is
expected that the structure of the tip density of states should be
reflected in STS results and can significant change the position
or intensity of the surface DOS peaks, \cite{Pelz}. Moreover, new
maxima can appear on the differential conductance which are not
connected with the surface DOS, e.g. \cite{Passoni}.  It is also
known that metal tips induce a band bending on the semiconductor
surface (tip-induced band bending effect), e.g.
\cite{Loth,Narita,Dombrowski}, which strongly influences the
tunnelling current and leads to new peaks in STS results.

In many STM experiments on single atoms, molecules or even flat
atomic surfaces the STS results (differential conductance curves)
obtained for positive and negative STM voltages are asymmetrical.
Also the topography images for both voltages are often quite
different, cf. \cite{Narita,Jal01}. The main reason of such
behavior is asymmetry in the density of states of investigated
objects. Such an object can be characterized itself by
asymmetrical DOS (e.g. highly asymmetric atomic structure in the
tunnelling current between neighboring graphite atoms was observed
in Ref. \cite{Tomanek}) or this asymmetry can appear e.g. due to
the coupling object-surface, \cite{Jal02}.  Moreover, non ideal
tip geometry can also lead to different pictures for positive and
negative voltages. Note, that in real STM experiments it is almost
impossible to obtain fully symmetrical results.

From the basic electrodynamic rules, one expects large charge
concentration on sharp or curved conducting materials, like e.g.
on STM tips. Charge occupation of such tips is very important as
concerns electrical properties and can lead to asymmetry in the
differential conductance (versus the positive and negative STM
voltages)  even for objects which are characterized by fully
symmetrical density of states. We show in this paper that the
sharper tip, the more charge is accumulated on it and the
conductance symmetry can be broken in this case. (Note, that this
effect is not relevant to electrostatic forces which appear due to
differences between the Fermi levels of a sample and the tip or
due to applied sample voltage, e.g. \cite{Dombrowski,Hofer}). The
goal of this paper is to show that additional charge localized at
the STM tip, in addition to other possible sources, influences the
current-voltage characteristics and is responsible for the
asymmetry effect in the STS studies. In our calculations we
consider three models of the STM tip and use the Green function
formalism together with a tight-binding Hamiltonian to obtain the
conductance through quantum states on a surface. To illustrate
experimentally our theoretical predictions the STS studies of thin
Ag films on Si(111)-(6$\times$6)Au substrate were performed.
During the measurement a sudden variation of the tip length took
place which allowed us to compare the STS results for two kinds of
STM tips (sharp and blunt). Note, that small changes in the tip
structure often is observed during a series of scans, e.g.
\cite{Pelz}. These changes have only a minor effect on topographic
pictures but can drastically alter the conductance curves, cf.
also \cite{Kra0}.

The paper is organized as follows. Additional charge at the STM
tip is analyzed in Sec.~\ref{sec30}. In Sec. \ref{sec20} the
Hamiltonian for three model tips and general formulas for the STS
current and the transmittance are shown. In Sec.~\ref{sec40} the
results for a single atom and short wire on a surface are
discussed. The STM experimental results and their theoretical
description are shown in Sec.~\ref{sec50}. The last section,
Sec.~\ref{sec60}, is devoted to  conclusions.

\section{\label{sec30}Charged STM tip}

The main idea of this paper is  that the STM tip is charged i.e.
it possesses an additional charge. This charge influences the STS
results and leads to asymmetry in the conductance or
current-voltage curves. In this section we analyze a simple model
of the STM tip and explain why an additional charge is localized
at the last tip atom.

\begin{figure}[tb]
\begin{center}
\includegraphics[width=0.45\columnwidth]{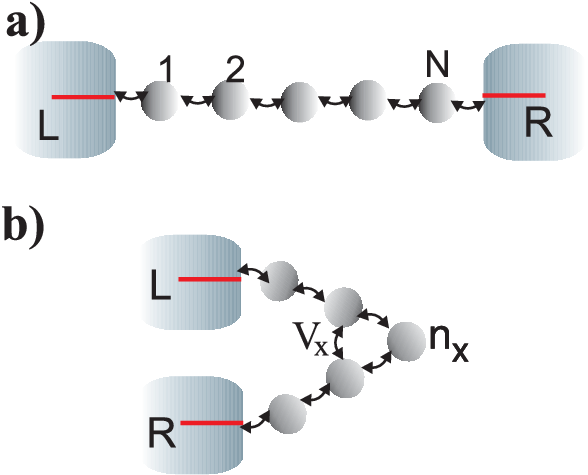}
\end{center}
\caption{The schematic view of (a) $N$-atom straight wire coupled
with the left and right electrodes, and (b) bend wire with nonzero
hopping, $V_x$, between $x-1$ and $x+1$ atoms in the wire. Figure
(b) represents the STM tip which is fabricated by bending a
straight one-dimensional wire (panel a). The figure depicts only a
part of junction, namely the tip without a sample. } \label{fig02}
\end{figure}
First, let us analyze the position (on the energy scale) of the
apex atom energy level, $\varepsilon_{STM}$, versus the chemical
potential of the STM electrode. For $\varepsilon_{STM} >
\mu_{STM}$ ($\varepsilon_{STM} < \mu_{STM}$) the tip is not
charged (is charged) and for $\varepsilon_{STM} = \mu_{STM}$ is
neutral. In most of our calculations we set the energy level of
the apex atom below the Fermi energy of the STM electrode because
in this case the STM tip is charged. Here we give the reason of
our choice. Using the basic electrodynamic relations one can show
that additional charge is gathered on sharp edges and curvatures.
To confirm this rule for the STM we consider here a simple model
of the STM tip i.e. one-dimensional bend monoatomic wire. In
Fig.~\ref{fig02}a the schematic view of a straight wire with the
nearest neighbor couplings between atoms, $V_0$, is shown. For a
bend wire, Fig.~\ref{fig02}b, also the next neighbor hopping has
to be considered, $V_x$. The Hamiltonian for this system is given
by
\begin{eqnarray}
H &=& \sum_{i=1}^N \varepsilon_{i} c^+_{i} c_{i}+\sum_{\vec k \in
{L/R}} \varepsilon_{\vec k} c^+_{\vec k} c_{\vec k} \nonumber\\
&+& \sum_{\vec k \in {L/R}} V_{\vec k} c^+_{\vec k} c_{1/N}
+\sum_{i=1}^{N-1} V_0 c^+_{i} c_{i+1}+V_x c^+_{x-1}c_{x+1} + h.c.
\,, \label{Hwire}
\end{eqnarray}
where $x$ stands for the middle atom in our wire, $x=(N+1)/2$, $N$
is odd. The main quantity of our interest is the occupation of
this middle atom, $n_x$, which represents the last tip atom. The
local charge at this atom can be expressed by means of the
retarded Green's function, $G^r_{xx}$ i.e. $n_x={-1 \over \pi}
\int_{-\infty}^{E_F} Im G^r_{xx}(\varepsilon) d\varepsilon$. The
function $G^r_{xx}$ can be obtained from the equation of motion
for the retarded Green's function and in our case [for $N$ atom
wire with the same electron energies,
$\varepsilon_i=\varepsilon_0$, and within the wide band
approximation, $\Gamma=\Gamma^L=\Gamma^R=2 \pi \sum_{\vec k\in L}
|V_{\vec k}|^2 \delta(\varepsilon-\varepsilon_{\vec k})$] it
becomes:
\begin{eqnarray}
G^r_{xx}(\varepsilon)={B_{x-1}^2-V_x^2B_{x-2}^2 \over B_N+i{\Gamma
\over 2} B_{N-1}-V_x[2V_0^2+V_x(\varepsilon-\varepsilon_0)]
B_{x-2}^2 } \label{FG}
\end{eqnarray}
where $B_x=\det A_x^0+i{\Gamma \over 2} \det A_{x-1}^0$, and
$A_x^0$ matrix  reads:
\begin{eqnarray}
A_{x}^0=\left(%
\begin{array}{ccccc}
  \varepsilon-\varepsilon_0 & -V_0 & 0 &   &   \\
  -V_0 & \varepsilon-\varepsilon_0 & -V_0 &   &   \\
  0 & -V_0 & \varepsilon-\varepsilon_0 &   &   \\
    &   &   &  \ddots &   \\
    &   &   & -V_0 & \varepsilon-\varepsilon_0 \\
\end{array}%
\right)_{x\times x} \label{matrix}
\end{eqnarray}
The determinant of this matrix can be expressed by means of the
Chebyshev polynomials of the second kind, \cite{Kwa01}, and
finally the retarded Green's function, $G_{xx}^r$, can be obtained
analytically. It helps us to find the occupation of the middle
atom in the wire, as a function of $V_x$ parameter.

\begin{figure}[tb]
\begin{center}
\includegraphics[width=0.5\columnwidth]{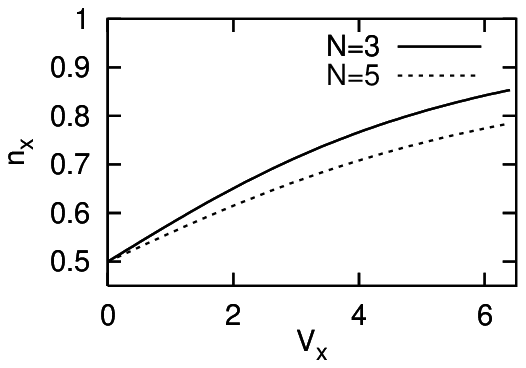}
\end{center}
\caption{The occupation of the middle atom in the wire as a
function of $V_x$ parameter (see Fig.~\ref{fig02}b) for $N=3$
(solid line) and $N=5$ (broken line). The other parameters are:
$\Gamma=1$, $\varepsilon_0=0$, $E_F=0$, $V_0=4$, $T=0$K (all
energies in $\Gamma$ units). } \label{fig03}
\end{figure}
In Fig.~\ref{fig03} we show the charge localized at the middle
atom versus the coupling parameter $V_x$ for the wire length $N=3$
(solid line) and $N=5$ (broken line). The parameter $V_x$ is
responsible for the wire curvature i.e. for $V_x=0$ the wire is
straight and for nonzero $V_x$ we have a kind of bend wire as is
depicted in Fig.~\ref{fig02}b.  Here we consider the case
$\varepsilon_0=\mu_L=\mu_R=0$ which leads to half occupation of
all wire sites for $V_x=0$ ($\mu_{L/R}$ is the chemical potential
of the $L/R$ electrode). However, for nonzero $V_x$ the occupation
of the middle atom increases and e.g. for $V_x=V_0=4$ it is about
$30\%$ larger than for the straight wire. This result indicates
unquestionably that there is additional electron charge at the
middle atom. This atom corresponds to the last tip atom and is
more occupied in comparison with other atoms.

Here we should comment on the last result. Of course, STM tips are
not fabricated by bending a straight monoatomic wire. However, a
sharp STM tip, with only one atom, can be described as a kind of
short bend wire e.g. with $N=3$ or $N=5$ atoms. In this case  the
coupling $V_x$ is responsible for the tip curvature and the larger
$V_x$ the sharper tip is (there is physical limit on $V_x$ - it
cannot be larger than twice or a few values of $V_0$ parameter).
Thus, this approach is suitable for description of sharp tips and
failed for rather wide ones. Note, that in the literature  STM
tips are often much more simplified and modelled by a
semi-infinite chain of atoms, e.g. \cite{Ou-Yang,McKinnon} or an
ideal fermi gas electrode (energy structure-less).

On the other hand, one can consider the last tip atom as a kind of
additional atom (adatom) which is put on metallic STM electrode.
This atom cannot be treated in the same way as atoms inside the
STM electrode, first of all because its neighborhood is different
in comparison with other atoms in the tip. Moreover, the energy of
affinity (and ionization) level at the apex atom depends on the
distance between this atom and the STM metallic surface. This
effect is known in chemisorption processes, i.e. the affinity
energy level decreases its value versus the surface Fermi energy
with the distance adatom-surface and is minimal for an adatom
placed directly on a surface, cf. \cite{Desjonq}. In that case
electrons from the surface can occupy the adatom affinity state
which leads to minimalization of the total energy (cf. also the
calculated local DOS of Co atoms at Au surface which is
characterized by local peaks under the Fermi level, \cite{Ujsa}).
This explanation support our idea that the STM tip is charged.
Also dynamical polarization effects (surface-adatom or
surface-molecule) can renormalize molecular states,
\cite{Thygesen}.

Taking into account the above considerations we assume in our
calculation that the energy of the last tip atom,
$\varepsilon_{STM}$, lays slightly below the STM chemical
potential, i.e. $\varepsilon_{STM} < \mu_{STM}$, and thus is more
occupied. One can obtain the same effect for $\varepsilon_{STM} =
\mu_{STM}$ but the STM tip should be considered as a bend wire,
cf. Fig.~\ref{fig02}b. In the last case the middle atom in the
wire is extra occupied (its density of states possesses a local
maximum below the Fermi energy). It allows us to treat the STM tip
as a metallic electrode with only one atom under the condition
$\varepsilon_{STM} < \mu_{STM}$.


\section{\label{sec20}Theoretical STM model}

\begin{figure}[tb]
\begin{center}
\includegraphics[width=0.8\columnwidth]{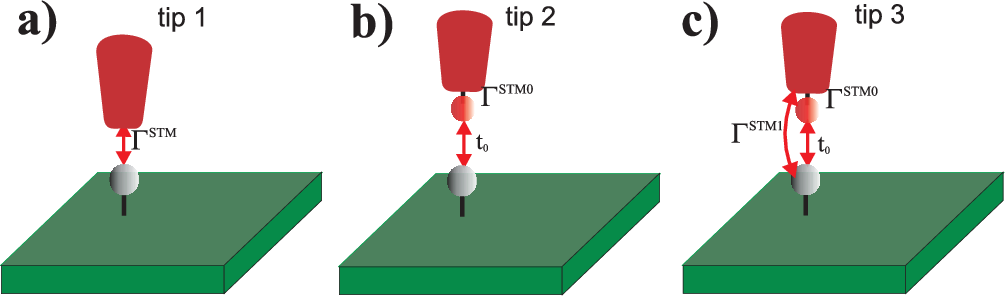}
\end{center}
\caption{Sketch of one-atom object on a surface and three STM tip
models: (a) STM tip as a metallic electrode, (b) tip consisted of
a metallic electrode with one atom (coupled with the STM electrode
via $\Gamma^{STM0}$); the apex atom is also coupled with the
investigated object via $t_0$ parameter, (c) the same as in (b)
but for nonzero coupling between the tip electrode and the object
via $\Gamma^{STM1}$ parameter. } \label{fig00}
\end{figure}

The system consists of the STM tip, an object (few-atom system,
i.e. one atom or short wire) and a metallic surface, cf.
Fig.~\ref{fig00} for only one atom on the surface. The STM tip can
be represented by (i) a metallic electrode i.e. tip 1,
Fig.~\ref{fig00}a - electrons flow from this electrode directly to
the investigated object, (ii) a metallic electrode with the last
tip atom on it, i.e. tip 2, Fig.~\ref{fig00}b - electrons flow
from the electrode through this apex atom then through the object
on the surface, and (iii) a metallic electrode with the apex atom
but electrons can flow from the STM electrode directly to the
object or through the last tip atom - tip 3, Fig.~\ref{fig00}c.
Tip 2 in Fig.~\ref{fig00} is suitable for very sharp STM tips and
tip 1 (widely used in the literature) describes rather blunt tips.
The third case, tip 3, can be treated as a composition of the
first and the second tips. To obtain the current flowing from the
STM electrode we use the Green function formalism and
thigh-binding Hamiltonian, cf. \cite{Jal01,Hofer}, which can be
written as follows:
\begin{equation}
H =H_{STM}+H_{surf}+H_{object}+\tilde{V} \,, \label{H1}
\end{equation}
where $H_{STM}$ describes electrons in the STM tip. For a metallic
STM electrode it can be expressed in the form (ideal Fermi gas):
$H_{STM}= \sum_{\vec k_{STM}} \varepsilon_{\vec k_{STM}} c^+_{\vec
k_{STM}} c_{\vec k_{STM}}$, which corresponds to Fig.~\ref{fig00}a
or with one tip atom, cf. Fig.~\ref{fig00}b,c:
\begin{equation}
H_{STM}= \sum_{\vec k \in {STM}} \left( \varepsilon_{\vec k}
c^+_{\vec k} c_{\vec k}+t_{\vec k} c^+_{\vec k} c_{STM}+
h.c.\right) + \varepsilon_{STM} c^+_{STM} c_{STM}  \,, \label{H2}
\end{equation}
Here $\varepsilon_{\vec k_{STM}}$ and $\varepsilon_{STM}$
represent electron energies in the STM electrode and at the tip
atom. $c^+$ and $c$ are the creation and annihilation electron
operators in an appropriate electron state and $t_{\vec k}$ stands
for the coupling parameter between the STM electrode and the tip
atom ($\Gamma^{STM0}$ in Fig.~\ref{fig00} depends on this
parameter). The surface, similarly to the STM electrode,  is
modelled as an ideal Fermi gas by the Hamiltonian, $H_{surf}=
\sum_{\vec k_{surf}} \varepsilon_{\vec k_{surf}} c^+_{\vec
k_{surf}} c_{\vec k_{surf}}$. The object Hamiltonian depends on
the number of atoms at the surface and e.g. for a single atom it
can be written as $H_{object}=\varepsilon_{1} c^+_{1} c_{1}$, but
for a short wire it has the following form
\begin{equation}
H_{object}=\sum_{i=1}^N \varepsilon_{i} c^+_{i} c_{i} +
\sum_{i=1}^{N-1}  V_i c^+_{i} c_{i+1} +h.c. \, \label{H4}
\end{equation}
The coupling Hamiltonian is responsible for electron tunnelling
between the STM and the surface electrode and for $H_{STM}$,
Eq.~\ref{H2}, and $H_{object}=\varepsilon_{1} c^+_{1} c_{1}$, (tip
2, Fig.~\ref{fig00}b) it can be expressed as:
$\tilde{V}=\sum_{\vec k_{surf}} V_{\vec k_{surf}} c^+_{\vec
k_{surf}} c_{1} + t_0 c_1^+ c_{STM} +h.c.$, where $t_0$ is the
hopping integral between the tip atom and the object atom. Note,
that for $N$-site wire on the surface one should change the
coupling Hamiltonian and instead of $c_1$ operator one has to sum
$c_i$ operators over all atomic sites in the wire i.e.
$i=1,...,N$. Moreover, for the system shown in Fig.~\ref{fig00}c
one has to add to the Hamiltonian the term which is responsible
for the coupling between the STM electrode and the object atom.
Thus the coupling Hamiltonian becomes:
\begin{equation}
\tilde{V}=\sum_{i=1}^N \sum_{\vec k_{surf}} V_{i,\vec k_{surf}}
c^+_{\vec k_{surf}} c_{i} + t_0 c_1^+ c_{STM}+ \sum_{\vec k_{STM}}
V_{\vec k_{STM}} c^+_{\vec k_{STM}} c_{1} +h.c. \,, \label{H6}
\end{equation}
where we assume that the STM electrode is connected with the first
atom in the wire via $V_{\vec k_{STM}}$ element (it is responsible
for $\Gamma^{STM1}$ parameter, cf. Fig.~\ref{fig00}c). The role of
electron-electron interactions in the system is discussed in
Sec.~\ref{sec41b}. Electron transport properties are analyzed
within the framework of the Green's function method. The
tunnelling current flowing from the STM electrode to the surface
can be obtained from, \cite{Wingreen}:
\begin{equation}
I={2e\over h}\int d\varepsilon T(\varepsilon)
[f_{STM}(\varepsilon)-f_{surf}(\varepsilon+V)] \,, \label{curr}
\end{equation}
where $f(\varepsilon)$ is the Fermi function and the external
voltage is expressed by means of the STM and surface chemical
potentials i.e. $V=\mu_{STM} - \mu_{surf}$. The transmission
function for considered here systems reads:
\begin{equation}
T(\varepsilon)=Tr\{\hat{\Gamma}_{STM} \hat{G}^r
\hat{\Gamma}_{surf} \hat{G}^a\} \,, \label{trans}
\end{equation}
where $\hat{\Gamma}_{STM/surf}$ is the matrix composed of the
coupling parameters between the STM electrode or the surface with
the object atoms. In general the elements of these matrixes can be
written in the form $(\hat{\Gamma}_{surf})_{ij}=2 \pi \sum_{\vec
k_{surf}} V_{i,\vec k_{surf}}V^*_{j,\vec
k_{surf}}\delta(\varepsilon-\varepsilon_{\vec k_{surf}})$, and
similar for $\hat{\Gamma}_{STM}$. The next matrixes in
Eq.~\ref{trans} are composed of the retarded and advanced Green
functions ($\hat{G}^a=\hat{G}^{r*}$), and can be obtained from the
equation of motion for the retarded Green's function. Note, that
for  our systems, Fig.~\ref{fig00}, the transmittance can be
obtained analytically and all matrix elements will be specified
later. The knowledge of the current flowing through the system and
the STM voltage is sufficient to find the differential
conductance, $dI/dV$, or the normalized differential conductance,
$d(\ln I)/d( \ln V)$ .

\section{\label{sec40}Results and discussion}

In this section we show and analyze the current and differential
conductance as a function of the surface chemical potential
$\mu_{surf}$ ($\mu_{STM}=0$). It is worth noting that all object
atoms are placed directly on the surface and thus we assume that
their onsite energies are shifted with the surface chemical
potential, $\varepsilon_i \rightarrow \varepsilon_i+\mu_{surf}$.
The same is true for the last tip atom and the STM electrode. This
procedure is justify in our system as the largest potential drop
is observed for the weakest coupling - in our case this is $t_0$
parameter which is at least $10$ times smaller than the other
couplings (the tip and the substrate are in local equilibrium).
All results in this section are obtained for the energy unit
$\Gamma=1$. The current and conductance are expressed in the units
of $2e\Gamma/h$ and $2e^2/h$, respectively, and e.g. for
$\Gamma=10^{-3}$eV the current unit corresponds to $10^{-8}$A. The
other parameters have been chosen in order to satisfy the
realistic situation in many STM experiments, e.g.
\cite{Jal01,Jal02,Kra0}.

Note that in our calculations we chose fully symmetrical objects
in the energy space. The first object is a single atom on a
surface which is characterized by the Lorentz-type density of
states (DOS). The second one stands for a short wire also with
symmetrical DOS. It is obvious that in real situation such atomic
objects on a surface can be characterized by asymmetrical DOS e.g.
due to different couplings object-surface. In that case the
differential conductance curves have to be also asymmetrical and
it is difficult to analyze the role of charged tip in this
asymmetry. To avoid this problem we concentrate here on objects
with symmetrical DOS.

\subsection{\label{sec41} STM tunnelling through a single atom}

For a single atom on the surface the following analytical
solutions for $\hat{\Gamma}$ and $\hat{G}^r$ matrixes can be
obtained:
\begin{itemize}
    \item Tip 1, Fig.~\ref{fig00}a:
    $\hat{\Gamma}_{STM}=\Gamma^{STM0}$, $\hat{\Gamma}_{surf}=\Gamma^{surf}$ and
    $\hat{G}^{r}=G_{11}^r=\left( \varepsilon-\varepsilon_1+i{\Gamma^{STM0}+\Gamma^{surf} \over 2}
    \right)^{-1}$, where $\Gamma^{STM0}=2 \pi \sum_{\vec k\in STM} |t_{\vec k}|^2
\delta(\varepsilon-\varepsilon_{\vec k})$, $\Gamma^{surf}=2 \pi
\sum_{\vec k\in surf} |V_{\vec k}|^2
\delta(\varepsilon-\varepsilon_{\vec k})$ are energy independent
(wide band limit approximation). According to the above relations
the transmittance becomes:
\begin{eqnarray}
T(\varepsilon)={\Gamma^{STM0} \Gamma^{surf} \over  \left(
\varepsilon-\varepsilon_1\right)^2+\left({\Gamma^{STM0}+\Gamma^{surf}
\over 2} \right)^2  }
 \,
\end{eqnarray}
    \item Tip 2, Fig.~\ref{fig00}b (there are two atoms and we assume that the index
    1/2 in all matrixes corresponds to the object/STM tip atom) nonzero matrix elements
    are: $(\hat{\Gamma}_{STM})_{22}=\Gamma^{STM0}$,
    $(\hat{\Gamma}_{surf})_{11}=\Gamma^{surf}$,
and the transmittance is expressed as follows:
$T(\varepsilon)=\Gamma^{STM0} \Gamma^{surf} |G^r_{12}|^2$, where
\begin{eqnarray}
 |G^r_{12}|^2&=&{ t_0^2
\over |\left( \varepsilon-\varepsilon_1+i{\Gamma^{surf} \over
2}\right) \left( \varepsilon-\varepsilon_{STM}+i{\Gamma^{STM0}
\over 2}\right)-t_0^2|^2  }
 \, \label{G12}
\end{eqnarray}

    \item Tip 3, Fig.~\ref{fig00}c, nonzero matrix elements are:
    $(\hat{\Gamma}_{STM})_{11}=\Gamma^{STM1}$,
    $(\hat{\Gamma}_{STM})_{22}=\Gamma^{STM0}$ and
    $(\hat{\Gamma}_{surf})_{11}=\Gamma^{surf}$
thus the transmittance becomes:
\begin{eqnarray}
T(\varepsilon)&=&\Gamma^{surf} \left( \Gamma^{STM0} |G^r_{12}|^2
+\Gamma^{STM1} |G^r_{11}|^2\right)
 \,,
\end{eqnarray}
where
\begin{eqnarray}
 |G^r_{11}|^2&=&{ |\varepsilon-\varepsilon_{STM}|^2
\over |\left(
\varepsilon-\varepsilon_1+i{\Gamma^{surf}+\Gamma^{STM1} \over
2}\right) \left( \varepsilon-\varepsilon_{STM}+i{\Gamma^{STM0}
\over 2}\right)-t_0^2|^2  }
 \, \label{G122}
\end{eqnarray}
\begin{eqnarray}
 |G^r_{12}|^2&=&{ t_0^2
\over |\left(
\varepsilon-\varepsilon_1+i{\Gamma^{surf}+\Gamma^{STM1} \over
2}\right) \left( \varepsilon-\varepsilon_{STM}+i{\Gamma^{STM0}
\over 2}\right)-t_0^2|^2  }
 \, \label{G121}
\end{eqnarray}

\end{itemize}
The knowledge of the transmittance allows us to find the current
flowing in the system and the differential conductance.

\begin{figure}[tb]
\begin{center}
\includegraphics[width=0.45\columnwidth]{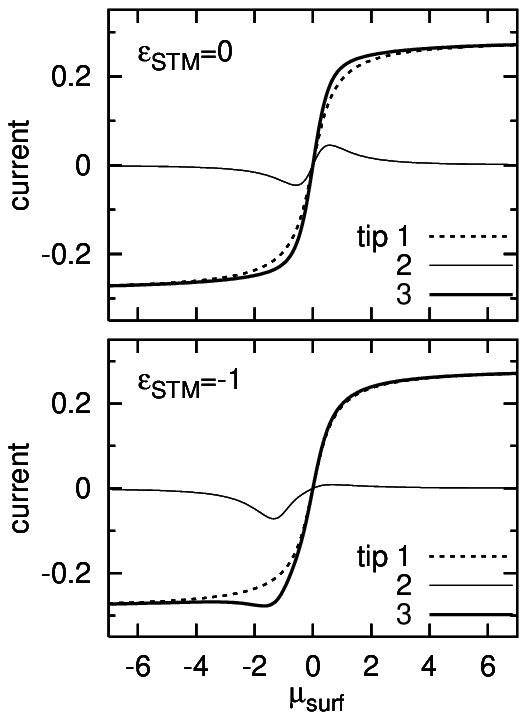}
\end{center}
\caption{Current-voltage characteristics for the neutral
($\varepsilon_{STM}=0$, upper panel) and occupied STM atom
($\varepsilon_{STM}=-1$, lower panel). The broken (thin solid,
thick solid) lines correspond to tip 1 (tip 2, tip 3)
configuration shown in Fig.~\ref{fig00}. The other parameters are:
$\Gamma^{STM0}=\Gamma^{surf}=\Gamma=1$, $\Gamma^{STM1}=0.1$,
$t_0=0.1$, and $\varepsilon_1=\mu_{surf}$, $\mu_{STM}=0$.}
\label{fig04}
\end{figure}
In order to reveal the role of charged tip we investigate the STM
current for all described tip models. In Fig.~\ref{fig04} the
current-voltage characteristics are shown for the case of the
neutral tip (upper panel) and the occupied one (lower panel). The
broken lines are obtained for the simplest model where the STM tip
is represented by a metallic electrode without any tip atom, cf.
Fig.~\ref{fig00}a. Thus the broken line is independent on
$\varepsilon_{STM}$ parameter and is the same in both panels. The
current in this case is monotonic function of the voltage. The
thin solid (thick solid) lines correspond to tip 2,
Fig.~\ref{fig00}b (tip 3, Fig.~\ref{fig00}c). The currents
obtained within both tip 1 and tip 3 are very similar because the
object atom for these cases is coupled directly with the STM
electrode, cf. the broken and thick lines. In contrast to this,
the current is not monotonic function for tip 2 and takes nonzero
values only for $\mu_{surf} \approx \varepsilon_{STM}$, cf. thin
solid lines, both panels. In the last case electrons can tunnel
from the STM electrode to the surface only though the tip atom and
the object atom. This is the reason why the current flows if there
are nonzero local density of states at both atoms in the window of
the chemical potentials. This effect is also reflected in the
current curve obtained for tip 3 model for $\varepsilon_{STM}=-1$
(thick line, lower panel). In this case the current is
characterized by a local minimum for
$\mu_{surf}=\varepsilon_{STM}$. The current-voltage
characteristics obtained for the neutral and occupied tip atom
seem to be similar to each other, cf. the upper and the lower
panel for the broken, thick or thin lines, respectively. However,
there are very important and crucial for our paper differences
which become prominent on the differential conductance curves.

\begin{figure}[tb]
\begin{center}
\includegraphics[width=0.45\columnwidth]{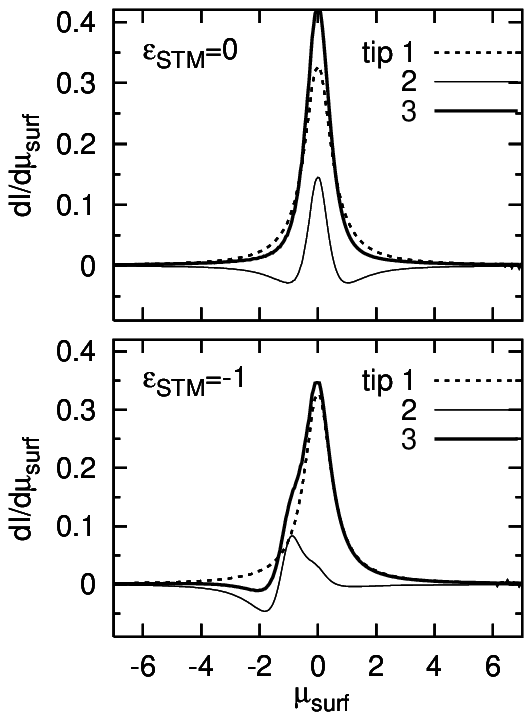}
\end{center}
\caption{Differential conductance, $dI/d\mu_{surf}$, obtained for
the currents from Fig.~\ref{fig04}. All parameters are the same as
in Fig.~\ref{fig04}. } \label{fig05}
\end{figure}
In Fig.~\ref{fig05} we show the differential conductance obtained
for the corresponding current curves from Fig.~\ref{fig04}. Note
that the conductance curves obtained according to tip 1 are the
same at both panels (cf. broken lines) and reflect the local
density od states of the object atom. As a most prominent feature,
the conductance turns out to be symmetrical versus $\mu_{surf}=0$
for the case of neutral STM tip (upper panel) and asymmetrical for
charged tip atom (lower panel). The explanation of this fact is as
follows, i.e. for the neutral tip the probability of electron
tunnelling  from the surface to the STM tip or vice versa is the
same (the absolute currents for positive and negative voltages are
the same, cf. Fig.~\ref{fig04}, upper panel). If the tip is
charged these probabilities are not equal and for electrons it is
easier to tunnel from the tip to the surface states than from the
surface to the STM electrode.

It is interesting that in the presence of the tip atom we find the
negative differential conductance, cf. thin solid lines in
Fig.~\ref{fig05} obtained for tip 2. The negative conductance is
related to the discrete structure of the local density of states
at both (STM and object) atoms (cf. \cite{Kra0} where the negative
conductance was observed for non-constant surface density of
states). If the overlap between the tip and the surface states
decreases (with increasing chemical potential) the negative
differential conductance appears, \cite{Xue}, (see \cite{Johan}
for other explanation of NDC). This condition can be satisfied
only for certain range of parameters.  For tip 3, thick lines, the
negative conductance appears only for the case of charged tip
(lower panel) and due to the direct coupling $\Gamma^{STM1}$, it
is not as prominent as for tip 2 ($\Gamma^{STM1}=0$, thin lines).
It is worth to comment on the last result. In real STM experiments
it is very difficult to observe the negative conductance mainly
due to not very sharp tip, temperature effects or direct coupling
between the STM and the surface electrode (not through the
object). The last factor gives only constant positive value to the
conductance (the corresponding current depends linear only on the
chemical potential difference) and shift the conductance curves
above the OX axis. Thus in our calculations we omit this coupling
and concentrate on the charged tip effects.

\begin{figure}[tb]
\begin{center}
\includegraphics[width=0.45\columnwidth]{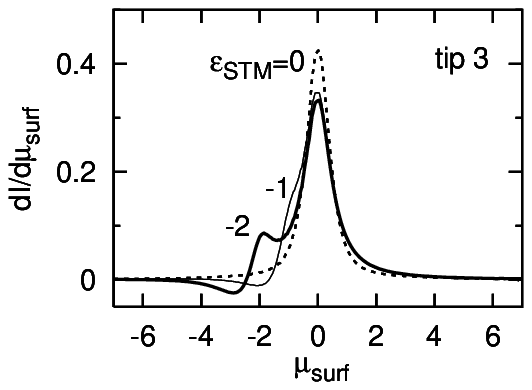}
\end{center}
\caption{Differential conductance, $dI/d\mu_{surf}$, obtained for
tip 3 and for $\varepsilon_{STM}=0 , -1, -2$ - broken, thin solid
and thick solid lines, respectively. All parameters are the same
as in Fig.~\ref{fig04}. } \label{fig06}
\end{figure}
The next intriguing question is whether the energy position of the
last tip atom, $\varepsilon_{STM}$, influences the asymmetry
effect. To investigate this problem we chose the third model of
the STM system, (viz tip 3, Fig.~\ref{fig00}c). In
Fig.~\ref{fig06} we show the differential conductance curves
obtained for $\varepsilon_{STM}=0$ (the neutral tip, broken line),
$\varepsilon_{STM}=-1$, and $\varepsilon_{STM}=-2$, thin solid and
thick solid lines, respectively. For $\varepsilon_{STM}<0$
(charged tip) we observe, as before, asymmetrical conductance
curves versus the zero voltage. Moreover, the more charged is the
tip the stronger asymmetry is observed. A common feature of all
curves is the emergence of peaks for
$\mu_{surf}=\varepsilon_{STM}$ and $\mu_{surf}=\varepsilon_1$,
which is visible especially for the thick curve,
$\varepsilon_{STM}=-2$ (the peak for $\mu_{surf}=-2$ is related to
the energy level of the apex STM atom). This effect could be used
to test the quality of the STM tip. The apex tip atom leaves its
fingerprints in the differential conductance curves: First, the
conductance curves are asymmetrical and second, the conductance
peak related to the tip atom should appear. Note, that using
first-principles calculations, the studies of the geometrical,
electronic, and dynamic properties of a single atoms adsorbed on
silicon or tungsten surfaces were reported e.g. in Ref.
\cite{Chen,Hofer0}.

\subsection{\label{sec41b} STM tunnelling through a single atom - the role of Coulomb interaction}

In this subsection we analyze the influence of Coulomb interaction
between the STM tip and one-atom object on the electron transport
through this system, cf. Fig.~\ref{fig00}c. The interaction
Hamiltonian can be written in the following form:
\begin{equation}
H_{int}=U c^+_{1} c_{1} c^+_{STM} c_{STM} \, \label{U1}
\end{equation}
Here we restrict our investigation to non-magnetic atoms and do
not consider many-body effects like the Kondo effect, charge-spin
separation effect and others. It allows us to use a Hartree-Fock
like approximation and the interaction can be captured by
renormalizing the on-site energy levels. In this case we can
substitute $\varepsilon_1$ by $\varepsilon_1+{U\over 2}n_{STM}$
and $\varepsilon_{STM}$ by $\varepsilon_{STM}+{U\over 2}n_{1}$
where $n_{1/STM}$ stands for the charge occupation of the atom on
a surface and the STM apex atom, respectively. The occupations,
$n_{1/STM}$, can be obtained from the the knowledge of the
retarded Green functions according to the relation $n_1={-1 \over
\pi} \int_{-\infty}^{\mu_{surf}} Im G^r_{11}(\varepsilon)
d\varepsilon$, and similar for $n_{STM}$. In this case it is
impossible to derive analytical solutions for the current or the
conductance as the retarded Green function depends on both (the
surface and the STM) atom occupations and is found numerically in
the self-consistent way.

\begin{figure}[tb]
\begin{center}
\includegraphics[width=0.45\columnwidth]{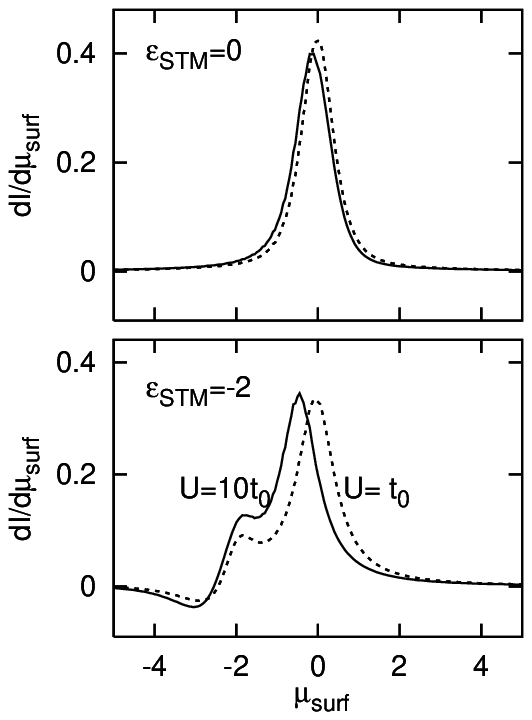}
\end{center}
\caption{Differential conductance for the neutral
($\varepsilon_{STM}=0$, upper panel) and occupied tip atom
($\varepsilon_{STM}=-2$, lower panel). The broken (solid) lines
correspond to the Coulomb interaction $U=t_0$ ($U=10t_0$). Tip 3
is considered and the other parameters are the same as in
Fig.~\ref{fig06}. } \label{figU}
\end{figure}
In Fig.~\ref{figU} the influence of Coulomb interaction on the
differential conductance through one-atom object on a surface is
studied. The upper (lower) panel corresponds to the neutral,
$\varepsilon_{STM}=0$ (occupied, $\varepsilon_{STM}=-2$) STM tip.
For very week interaction, $U=t_0$, the differential conductance
curves are very similar to those ones obtained for $U=0$, cf. the
broken and thick solid lines in Fig.~\ref{fig06} with the broken
lines in Fig.~\ref{figU}. For stronger Coulomb interaction,
$U=10t_0$ there are subtle effects in the differential conductance
curves in comparison with the case for $U=0$. The main conductance
peak related to the surface atom state, for $\mu_{STM}=0$, is
slightly changed for the neutral STM tip. For the occupied STM tip
the position of this peak is shifted towards the negative values
of $\mu_{STM}$. It results from the renormalization of the on-site
atom energy i.e. the STM apex atom is strongly charged which
shifts the on-site surface atom energy above the fermi level -
this leads to the conductance peak below $\mu_{STM}=0$. The
results shown in Fig.~\ref{figU} confirm that the Coulomb
interactions slightly change the differential conductance
structure and can be omitted in our studies.

Moreover, the electron-electron Coulomb on-site interactions are
also neglected in our system as we consider here non-magnetic
atoms (e.g. a tungsten tip, an object made of silver, gold or lead
atoms and silicon-gold surface, cf. \cite{Jal01,Jal02}) and, as
was mentioned above, do not study many-body effects.

\subsection{\label{sec42} STM tunnelling to a short wire}

In order to corroborate the results of the previous subsection we
have computed the STM differential conductance for a quantum wire
on a surface (instead of the single atom considered in
Sec.\ref{sec41}). The wire consists of $N=3$ atoms with the same
onsite energies, $\varepsilon_1=\varepsilon_2=\varepsilon_3$, and
the nearest neighbor couplings, $V_i=V_0$. The coupling
wire-surface is described by the function
$\Gamma^{surf}_{ij}=\Gamma^{surf} \delta_{i,j}$, and
$\Gamma^{surf}$ is defined as before. The wire is characterized by
fully symmetrical local density of states: It is very important in
these studies - for asymmetrical DOS the conductance is also
asymmetrical. In our calculations we assume that the STM tip is
connected with the first wire atom (the results for other
connections are similar). The calculated method is the same as in
the previous subsection thus here we give only the relations for
the transmittance through three considered systems (cf.
Fig.~\ref{fig00} but for 3-site wire on the surface):
\begin{itemize}

    \item Tip 1, Fig.~\ref{fig00}a: $
T(\varepsilon)=\Gamma^{STM0} \Gamma^{surf}
\left(|G^r_{11}|^2+|G^r_{12}|^2+|G^r_{13}|^2 \right)$.

    \item Tip 2, Fig.~\ref{fig00}b (there are four atoms and we assume that
    indexes: 1,2,3 describe the wire sites and index 4 corresponds to the tip
    atom): $T(\varepsilon)=\Gamma^{STM0} \Gamma^{surf}
\left(|G^r_{14}|^2+|G^r_{24}|^2+|G^r_{34}|^2 \right)$.

    \item Tip 3, Fig.~\ref{fig00}c: $T(\varepsilon)=\Gamma^{STM0} \Gamma^{surf}
\left(|G^r_{14}|^2+|G^r_{24}|^2+|G^r_{34}|^2 \right) +\\
\Gamma^{STM1} \Gamma^{surf}
\left(|G^r_{11}|^2+|G^r_{12}|^2+|G^r_{13}|^2 \right)$.

\end{itemize}
and the retarded Green functions are obtained from the equation of
motion for these functions. Note, that the transmittance for tip 3
is not a simple sum of the transmittance obtained for tip 1 and
tip 2 due to $\Gamma^{STM1}$ parameter which appears directly in
the transmittance relation and also in the above Green functions.

\begin{figure}[tb]
\begin{center}
\includegraphics[width=0.45\columnwidth]{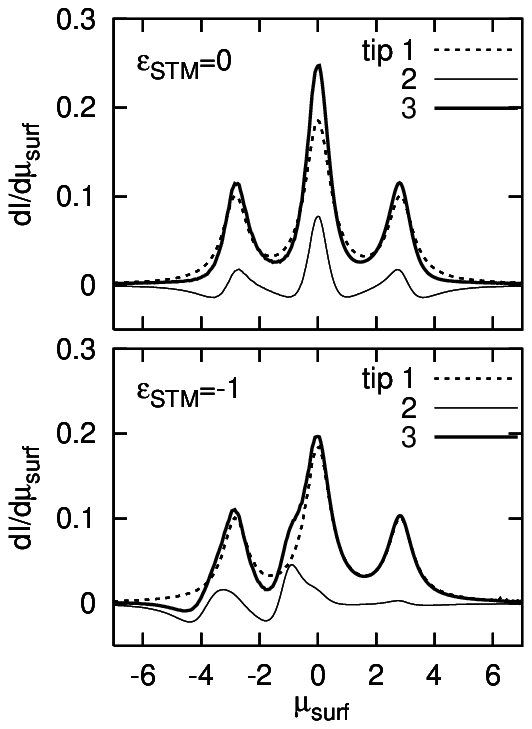}
\end{center}
\caption{Differential conductance for the neutral
($\varepsilon_{STM}=0$, upper panel) and occupied tip atom
($\varepsilon_{STM}=-1$, lower panel). The broken, thin solid and
thick solid lines correspond to tip 1, tip 2 and tip 3,
respectively. $V_0=2$, and the other parameters are the same as in
Fig.~\ref{fig04}. } \label{fig07}
\end{figure}
Figure~\ref{fig07} depicts the differential conductance obtained
for short quantum wire ($N=3$ atomic sites) for the neutral
($\varepsilon_{STM}=0$, upper panel) and occupied tip atom
($\varepsilon_{STM}=-1$, lower panel). Such a wire is
characterized by three-peak density of states which is reflected
also in the differential conductance curves (cf. the broken line
obtained for tip 1, upper panel). The thin solid (thick solid)
lines correspond to tip 2 (tip 3) and these peaks are also visible
in these cases. For the neutral tip all conductance curves are
fully symmetrical (upper panel), cf. also Fig.~\ref{fig05}.
However, for the occupied tip the conductance curves are
asymmetrical versus $\mu_{surf}=0$ (lower panel) which is in
accordance with the previous results. This asymmetry in the
conductance is very well visible for the thin solid line obtained
for tip 2 (Fig.~\ref{fig07}, lower panel, thin line). Here, for
$\mu_{surf}=-3$ a local conductance peak appears but for
$\mu_{surf}=3$ there is no such maximum in the differential
conductance. The conductance curve obtained for tip 3 (thick line)
is also asymmetrical. It is very interesting and important fact
that the occupied tip influences all conductance peaks (cf. the
thick and thin lines, lower panel) and not only the peak near
$\mu_{surf}=\varepsilon_{STM}$. For
$\mu_{surf}=\varepsilon_{STM}=-1$ one additional peak appears
which is the same effect as observed for the case of $N=1$,
Fig.~\ref{fig05}. This peak  is very well visible for tip 2 (thin
line) but for tip 3 the main conductance peak for $\mu_{surf}=0$
compensates the peak for $\mu_{surf}=-1$. Moreover, as before, the
negative differential conductance is observed for the results
obtained for tip 2 and tip 3.

\begin{figure}[tb]
\begin{center}
\includegraphics[width=0.55\columnwidth]{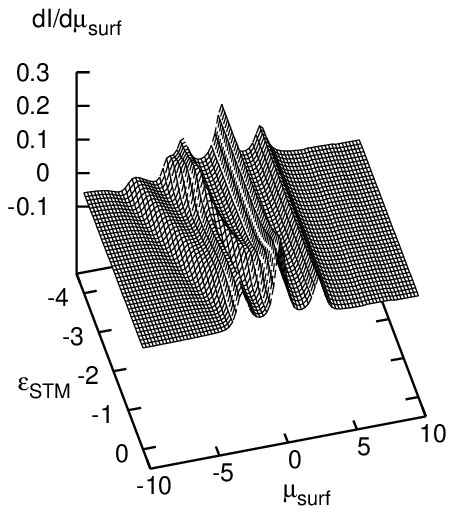}
\end{center}
\caption{Differential conductance as a function of the surface
chemical potential, $\mu_{surf}$, and the position of the last tip
atom, $\varepsilon_{STM}$ for the same parameters as in
Fig.~\ref{fig07} obtained for tip 3.} \label{fig08}
\end{figure}
A remaining question is whether the position of
$\varepsilon_{STM}$ influences in the same way all differential
conductance peaks. Therefore we plot in Fig.~\ref{fig08} the
differential conductance as a function of the surface chemical
potential, $\mu_{surf}$, and $\varepsilon_{STM}$. Here, the
symmetry of the conductance for the neutral tip is visible for
$\varepsilon_{STM}=0$ and asymmetrical shapes of $dI/dV$ for
negative $\varepsilon_{STM}$. It is interesting that the occupied
tip influences mainly the conductance for negative potentials i.e
for $\mu_{surf}<0$. For positive $\mu_{surf}$ the conductance peak
is slightly modified. Moreover, in Fig.~\ref{fig08} the evolution
of the additional peak which appears for
$\mu_{surf}=\varepsilon_{STM}$ is visible. For small values of
$\varepsilon_{STM}$ this peak is compensated by the main
conductance peaks (which appear for $\varepsilon_{STM}=0$), cf.
also Fig.~\ref{fig07}.

Of course, in the STM experiments it is very difficult (or even
impossible) to change the tip occupation or $\varepsilon_{STM}$
parameter. However, it is possible to use different kind of tips
(blunt or sharp) or other atoms attached to the tip. In the last
case during the measurement of the same object, one should obtain
different asymmetries of the conductance as each atom is
characterized by different $\varepsilon_{STM}$ parameter. This
experimental procedure, however, can destroy the tip or change its
geometry which can also lead to asymmetrical results.

\section{\label{sec50}Comparison with experiment}

\begin{figure}[tb]
\begin{center}
\includegraphics[width=0.5\columnwidth]{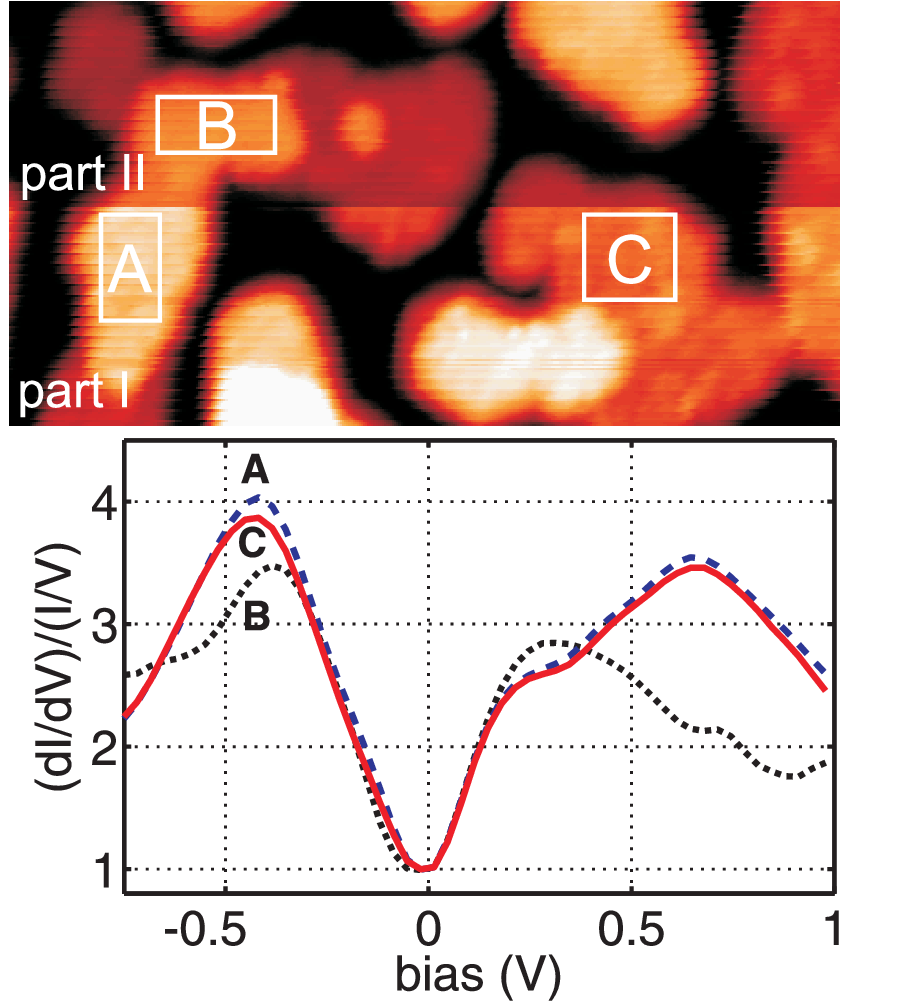}
\end{center}
\caption{(color online). Upper panel - $30\times15$ nm$^2$ STM
image of an Ag sample deposited onto Si(111)-$6\times6$ surface.
The rectangles A, B, and C mark areas for which  spectroscopic
$I(V)$ characteristics are analyzed. The scanning parameters are:
$V_s = -1.3 V$, $I = 1 nA$. Note image contrast difference between
part I and part II by sudden STM tip modification. Lower panel -
normalized $dI/dV$ curves of corresponding areas A, B, and C
marked in the upper panel. Curves A and B are recorded on the same
island, curves A and C are recorded on islands with different
thickness. Set point before opening the feedback-loop was $V_s =
-1.3 V$, $I = 1 nA$.} \label{fig001}
\end{figure}

In this subsection we compare our theoretical results with the STM
experiment on thin Ag films. Ag was grown on
Si(111)-(6$\times$6)Au substrate held at room temperature in an
UHV condition. The vacuum chamber was equipped with a RHEED
(Reflection High Electron Energy Diffraction) apparatus and STM
(type OMICRON VT). Electrochemically etched and annealed in UHV W
tip was used. The base pressure was $7\times10^{-11}$mbar. The
average thin Ag film thickness was equal to 3 ML of Ag(111). This
procedure yields flat Ag islands with (111) orientation, as
indicated by the RHEED pattern appearance. Figure \ref{fig001},
upper panel,  shows a constant-current scanning tunneling
microscope image of the sample which was used for current-voltage
spectra measurements. The sample was composed of well separated
islands with some hexagonal shape. Close inspection of the
island's surface morphology reveals presence of grainy features
with diameter of about $1.5$ nm. Consequently, the surface appears
rough, independent of the island height. Although RHEED patterns
obtained for this sample showed streaky features corresponding
epitaxially ordered structure, somewhat diffuse width of the
streaks indicated imperfect order. On Ag islands denoted as A, B,
and C in Fig.~\ref{fig001}, upper panel, were measured
current-voltage spectra and calculated the quantity
$(dI/dV)/(I/V)$, which is related to the density of states
\cite{Passoni2,Feenstra}. The thickness of islands determined from
the profile lines and counted from the wetting layer surface was
equal to 1.05 nm for areas A and B and 0.8 nm for the island C.
Figure~\ref{fig001}, lower panel, shows the STS spectra measured
on Ag islands shown in the upper panel of Fig.~\ref{fig001}. Each
spectrum in the lower panel is the average of 60-90 individual
$I(V)$ spectra in order to enhance the signal-to-noise ratio. The
experiment reported here is a good example of a phenomenon
frequently occurring during STM measurement practice - a sudden
variation of the tip length which is a consequence of the tip apex
rearrangement. Here the scan begins at the bottom, and in the
middle the average level of the image lowers of 1.5{\AA}.
Apparently, the tip became shorter. In order to keep the tunneling
current constant, the STM scanner approached the sample. It is
also visible that the resolution achieved for the lower part is
better than for the upper part of the image. Thus the tip switches
from "sharp" configuration (part I) to the "blunt" one (part II).
As the both areas A and B are on the same island, with the same
thickness, the electronic structure of the sample side of the
tunneling junction is identical. We note that in following
experiment on the same sample we observed also reversible
variations of the tip length. Figure~\ref{fig001} shows no
evidence of touching of the sample and one can assume that in
course of the scan the tip has lost the very top element,
presumably a single atom.

The variation of the topographic image is in accord with
electronic structure changes shown in Fig.~\ref{fig001}, lower
panel. The curve A, from the area scanned by the sharp tip (part
I), is displayed together with the curve B measured on the area B
with the blunt tip (part II). As a most prominent feature we
observe two quantum states at the surface voltages $V=-0.4$eV and
$V=+0.3$eV for both tips (curves A and B) and one additional state
for $V\simeq+0.7$eV observed only for the sharp tip (curve A). For
comparison we also measured spectra of the island C with the
thickness 1 ML smaller than for A and B. It is shown in
Fig.~\ref{fig001}, lower panel, as curve C. The curves A and C are
essentially identical. This fact excludes thickness-related
changes of the thin film electronic structure, namely the Quantum
Size Effect (QSE) \cite{Chiang}, as well as thickness-dependent
shift of the Ag surface state binding energy, as observed during
photoemission experiment \cite{Patthey}. Although the detailed
knowledge of the origin of the states observed in STS spectra is
not important for the following discussion, we expect that they
are relevant to the grainy structure of the island. Therefore, one
of the possible candidates are electronic states of the silver
clusters, like the clusters on graphite surface, studied with STS
in \cite{Hovel}.

\begin{figure}[tb]
\begin{center}
\includegraphics[width=0.45\columnwidth]{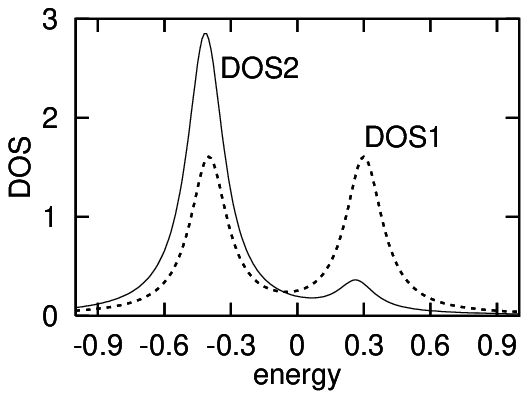}
\end{center}
\caption{Density of states obtained for two coupled atoms, $N=2$,
on a surface. The broken line corresponds to
$\varepsilon_1=\varepsilon_2=-0.05$, $V_0=0.35$ (symmetric case),
the solid line is obtained for $\varepsilon_1=-0.35$,
$\varepsilon_2=0.2$, $V_0=0.2$ (asymmetric case). } \label{figDOS}
\end{figure}

To analyze theoretically the experimental results we use the
method described in the previous section and consider two STM
tips: a metallic electrode, tip 1 (Fig.~\ref{fig00}a), and the tip
with one apex atom, tip 2 (Fig.~\ref{fig00}b). The first one
corresponds to the STM-tip from part II of our experiment whereas
the second one is responsible for part I (the tip is sharper). Two
Ag surface states which are visible in Fig.~\ref{fig001}, lower
panel, for $V=-0.4$eV and $V=+0.3$eV are generated in our model by
means of two coupled atomic sites situated on a metallic surface
(for our purposes the origin of these states is not important).
Fig.~\ref{figDOS} depicts the density of states (for two atoms on
the surface) for the symmetrical (DOS1) and asymmetrical (DOS2)
cases which are used in our calculations.

\begin{figure}[tb]
\begin{center}
\includegraphics[width=0.45\columnwidth]{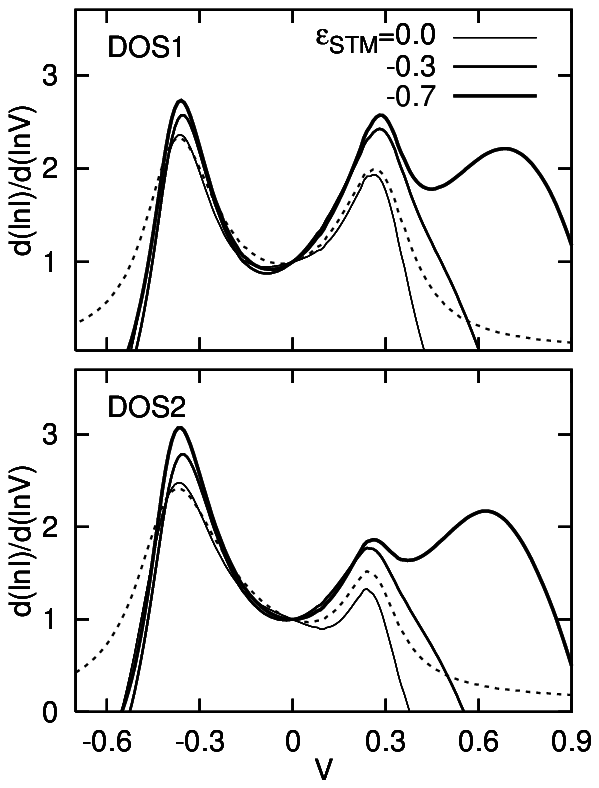}
\end{center}
\caption{Normalized differential conductance $d(\ln I)/d(\ln V)$
versus the surface voltage, $V=\mu_{STM}-\mu_{surf}$, obtained for
symmetrical DOS (upper panel) and asymmetrical DOS (lower panel)
and for different apex atom energy (according to tip 2 model):
$\varepsilon_{STM}=0, -0.3$ and $-0.7$ - thin solid, solid and
thick solid lines, respectively. Surface DOS are shown in
Fig.~\ref{figDOS}. The broken lines correspond to the metallic tip
without the apex atom (tip 1 model). The other parameters are:
$\Gamma^{STM0}=0.5$, $\Gamma^{STM1}=0$, $t_0=0.1$, and
$\mu_{STM}=0$. } \label{figpor}
\end{figure}
In Fig.~\ref{figpor} the normalized conductance as a function of
the surface voltage is shown for the symmetrical (upper panel) and
asymmetrical (lower panel) surface DOS (obtained according to tip
2 model) and for different tip occupation, $\varepsilon_{STM}=0,
-0.3$ and $-0.7$ - thin solid, solid and thick solid lines,
respectively. The case of $\varepsilon_{STM}=0$ corresponds to the
neutral tip and the results are very similar to the differential
conductance curves obtained for the tip without the apex atom i.e.
for tip 1, broken lines. Note, that for these cases (metallic tip
or the tip with neutral apex atom) the conductance curves are
characterized by two maxima which reflect only the surface quantum
states. This situation corresponds to the blunt STM tip used in
our experiment (part II). The thick solid lines in
Fig.~\ref{figpor} correspond to the case of occupied tip,
$\varepsilon_{STM}=-0.7$, and the additional peak for $V=+0.7$eV
is visible. (Remark: the STM tip states which are below the Fermi
level, on the conductance graphs are visible for positive surface
voltage). It is high probably that the origin of this peak is
related to the apex atom energy level which is below the Fermi
energy and is occupied (the surface states in the experiment are
the same for both parts but the tip has changed). The similar
effect appears in Fig. \ref{fig05} or Fig. \ref{fig06}, where for
charged tip additional structure in the conductance was observed.
Thus, the broken curves in Fig.~\ref{figpor} represent the STM tip
from part II and the thick solid lines correspond to the tip from
part I. Moreover, for asymmetrical DOS, lower panel, the
qualitative comparison of the differential conductance with the
experiment is better which suggests that the intensity of Ag
surface states (below and above the Fermi level) are different,
cf. Fig.~\ref{figDOS}, DOS2.

It is interesting that the normalized differential conductance
peaks are not proportional to the surface DOS and the results
obtained for symmetrical and asymmetrical DOS do not reflect
strong asymmetry which appears for DOS2, cf. Fig.~\ref{figDOS}. It
means that the intensity of $d(\ln I)/d(\ln V)$ peaks are not
directly proportional  to the surface DOS (the positions of these
peaks are related to the surface or STM DOS), cf. also
\cite{Passoni2}. Only for nearly constant tip density of states
the normalized differential conductance is proportional to the
sample DOS, \cite{Feenstra}. Thus, using STS results it is
difficult to distinguish between the surface and the STM tip
states due to their energy convolution. One possible way is to
scan two or more different areas of the surface (characterized by
different structure or different atoms) - if there are some states
which appear for both areas (for the same voltage) there are
probably the STM states.

To conclude, the STM states influence the conductance behavior and
disturb the intensity of the conductance peaks. For charged STM
tip, $\varepsilon_{STM}<0$, it leads to the symmetry braking in
the differential conductance, $dI/dV$, which is also visible on
the normalized differential conductance curves, $d(\ln I)/d(\ln
V)$. It is worth noting that the position of the additional peak
on the differential conductance curve, Fig.~\ref{figpor}, obtained
for $V=+0.7$eV is in good agreement with the results reported in
Ref.~\cite{Passoni}, Fig.~6, where the authors investigated
Au(111) surface by means of the tungsten tip and  observed a
special peak for $V=+0.6$eV which, in their opinion, was a halo of
the tip structure. In our studies the position of the tip-induced
peak is very similar which indicate that in experiments with sharp
tungsten tip  a local maximum around the sample voltage
$0.6-0.7$eV should be observed. This is also consistent with the
STS results on Cu(111) surface, \cite{Vazquez}, where the
differential conductance peak for the sample voltage
$V=0.5\pm0.3$eV appears and is identified as a halo of the
tungsten sharp STM tip (fcc pyramid configuration, cf. also
\cite{Hofer0}).

\section{\label{sec60}Conclusions}

In summary, using the Green's function technique for the
tight-binding Hamiltonian the STM transport properties through
surface states have been studied. It was shown that sharp STM tip,
represented by a kind of bending wire, is extra occupied. In this
case the analytical relation for the Green's function needed to
obtain the local charge has been obtained for arbitrary wire
length, $N$, Eq.~\ref{FG}.

The current and the differential conductance have been studied for
three models of the STM tip and for two objects on the surface
(single atom and short wire). The analytical formulas for the
transmittance and the retarded Green functions have been obtained.
For the neutral STM tip the conductance has turned out to be
symmetrical versus the surface chemical potential, (cf.
Fig.~\ref{fig05}, upper panel) and asymmetrical for charged tip
atom (Fig.~\ref{fig05}, lower panel and Fig.~\ref{fig08}). Thus
for occupied tip all differential conductance curves are
asymmetrical, although the density of states of investigated
objects are fully symmetrical in the energy space. Additional peak
in the conductance has been observed for the STM voltage which
corresponds to the onsite energy of the last tip atom. It means
that this peak is a feature of the STM tip and thus it should be
observed for different objects at the surface.

To confirm our theoretical calculations the STM experiment on Ag
islands with two different tips has been carried out. Additional
peak in the conductance curve for $+0.7$eV has been observed for
sharper tip which is a halo of the tip quantum state. This peak
has not been registered for the blunt tip, cf. Fig.~\ref{fig001}.
Our theoretical calculations obtained for charged STM tip,
Fig.~\ref{figpor}, are in good agreement with the experimental
results. Alternatively, such an experiment may  be also performed
with metallic molecules or clusters on a surface, e.g.
\cite{Gubin,Voss}. It has been also shown that the normalized
differential conductance peaks are not directly proportional to
the surface DOS. Moreover, the results obtained for symmetrical
and asymmetrical DOS do not reflect asymmetry in the peak
intensity of the surface DOS.

\vspace{1cm}
\noindent\textbf{Acknowledgements}. This work has been supported
by Grant No N N202 1468 33 of the Polish Ministry of Science and
Higher Education.


\bibliographystyle{elsarticle-num}
\bibliography{<your-bib-database>}



\section*{\label{sec4}Bibliography}

\end{document}